\documentclass[aps,prd,twocolumn,superscriptaddress,amsmath,amssymb,nobibnotes,nofootinbib]{revtex4-2}
\usepackage[utf8]{inputenc}
\usepackage[T1]{fontenc}
\usepackage{physics}
\usepackage{mathrsfs}
\usepackage{amsmath}
\usepackage{amsfonts}
\usepackage{amssymb}
\usepackage{braket}
\usepackage{graphicx}
\usepackage{enumitem}
\usepackage{cancel}
\usepackage{upgreek}
\usepackage[scr=boondox, cal=esstix]{mathalpha}
\usepackage[colorlinks=true,linkcolor=black,citecolor=black,urlcolor=black]{hyperref}

\begin{document}
\title{On the quantum mechanics of finite-mass observers}
\author{Juanca Carrasco-Martinez}
\affiliation{Leinweber Institute for Theoretical Physics, University of California, Berkeley, CA 94720, U.S.A.}
\affiliation{Theoretical Physics Group, Lawrence Berkeley National Laboratory, Berkeley, California 94720, USA}

\begin{abstract}
The principle of relativity is extended to accommodate finite-mass observers with quantum properties by introducing two operational requirements: (i) equivalence of observers at the level of transition amplitudes, and (ii) the impossibility for an observer to access its own quantum state of motion. This results in a fully relative formulation of quantum mechanics with observer-dependent Hilbert spaces, relative quantization rules, and novel uncertainty relations, while also elucidating some interpretational issues present in the current formulation of quantum mechanics and giving experimentally testable signatures.
\end{abstract}
\maketitle

\section{Introduction}
In classical and quantum mechanics, as presently understood, observers (reference frames or apparatus) are assumed to be free of disturbances during the measurement process, as they are considered ideal classical objects. While such an assumption is evident in the classical theory, since measurements do not play a central role, it persists in the quantum theory under the implicit condition that observers are infinitely massive and localized, so that their recoil and dynamical back-reaction can be neglected, effectively treating them as classical entities. 

When observers are instead described as quantum systems with arbitrary mass, nontrivial tensions emerge, such as the ability to detect the observer's proper state of motion. For example, in Ref.~\cite{Aharonov:1984zz}, to solve this apparent paradox, an additional vector potential is proposed, so that the relation between momentum and velocity is modified, $\hat p_\mathscr{k}^s \neq m \hat v_\mathscr{k}^s$, while maintaining $[\hat x_\mathscr{k}^s,\hat p_\mathscr{k}^s]=i\hbar$.

\begin{figure}[h!]
\vspace{-0.3cm}
    \centering
    \includegraphics[width=\columnwidth]{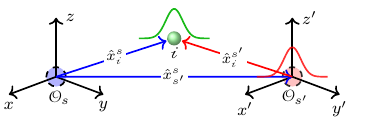}
\vspace{-0.5cm}    
    \caption{Representation of two observers $s$ and $s'$ measuring the particle $\mathscr{i}$.}
\label{Fig1}
\end{figure}

The origin of this tension can be made explicit in the setup shown in Fig.~\ref{Fig1}, where the following three standard conditions cannot be simultaneously satisfied:
\begin{enumerate}[leftmargin=11pt,label=\roman*.]
    \item For observer $s$, canonical quantization of $\mathscr{i}$ and $s'$ holds: $[\hat{x}_\mathscr{i}^s,\hat{p}_\mathscr{i}^s]
= [\hat{x}_{s'}^s,\hat{p}_{s'}^s] = i\hbar$.
    \item For observer $s'$, canonical quantization of $\mathscr{i}$ holds: $[\hat{x}_\mathscr{i}^{s'},\hat{p}_\mathscr{i}^{s'}] = i\hbar$.
    \item Conventional observable transformations: $\hat{x}^{s}_{\mathscr{i}} \equiv \hat{x}^{s}_{s'} + \hat{x}^{s'}_{\mathscr{i}}$, $\hat{p}^{s}_{\mathscr{i}} \equiv \hat{p}^{s'}_{\mathscr{i}} + \frac{m_\mathscr{i}}{m_{s'}}\hat{p}^{s}_{s'}$, where $m_\mathscr{i}$ and $m_{s'}$ are the masses of the particle $\mathscr{i}$ and observer $s'$, respectively.
\end{enumerate}
For instance, if conditions (iii) and (i) are imposed, then condition (ii) must be violated.

These three conditions are automatically satisfied in standard quantum mechanics since observers are treated as classical systems, and the operators $\hat{x}_{s'}^s$, $\hat{p}_{s'}^s$ effectively behave like c-numbers. As a consequence,  observables transform unitarily, and the canonical commutation relations are preserved in all reference frames as
\begin{equation*}
\begin{split}
&\hat{\mathcal{O}}_\mathscr{i}^{s'}={U}_s^{s'}\hat{\mathcal{O}}_\mathscr{i}^{s}{U}_s^{s'\dagger},\\
[\hat{x}_\mathscr{i}^{s}&,\hat{p}_\mathscr{i}^{s}]=[\hat{x}_\mathscr{i}^{s'},\hat{p}_\mathscr{i}^{s'}]=i\hbar.\\
\end{split}
\end{equation*}
Can we construct a quantum theory for finite-mass observers, given that the observable universe is itself finite, possibly quantum, and thus cannot operationally realize an ideal classical reference system?

Here, we answer in the affirmative by developing a formalism based on a relativity principle that is also valid for finite-mass observers (or quantum observers), which we refer to as the \emph{quantum relativity principle}.  Accordingly, the kinematics and quantum states are defined on observer-dependent Hilbert spaces without reference to any external classical description.

In the non-relativistic regime, the quantum relativity principle is implemented by transformations that preserve physical predictions and quantization rules that are explicitly relative. Concretely, in the setup of Fig.~\ref{Fig1}, the transformation law (iii) is retained while the canonical conditions (i) and (ii) are generalized as 
\begin{equation}
\begin{split}
[\hat{x}_\mathscr{i}^s, \hat{p}_\mathscr{j}^s]
&= i \hbar \hat{\mathcal A}^s_{\mathscr{i}\mathscr{j}},\quad[\hat{x}_{\mathscr{i}'}^{s'}, \hat{p}_{\mathscr{j}'}^{s'}]
= i \hbar \hat{\mathcal A}^{s'}_{\mathscr{i}'\mathscr{j}'},\\
\end{split}
\end{equation}
where $\hat{\mathcal A}^{(obs)}_{(\mathscr{a},\mathscr{b})}$ depends on both the observer and particles' $(\mathscr{a},\mathscr{b})$ properties, in contrast to prior proposals \cite{Aharonov:1984zz, Aharonov:1967zza, Giacomini:2017zju}.

Additionally, to satisfy the quantum principle of relativity, the new commutation relations are proven to be explicitly relational, non-canonical, and covariant under transformations between observers $\mathscr{T}_s^{s'}$, in which the classical description is recovered as a limiting case. Furthermore, incorporating the quantum relativity principle provides new perspectives on longstanding contrived interpretations in the conventional quantum theory, as discussed in the Wigner’s friend thought experiment.

Finally, we outline a test involving an observer (of mass $m_s$)  performing sequential measurements on two identical objects (of mass $m_\mathscr{i}$), where deviations from standard quantum mechanics arise at the order of $\mathscr{O}\big(\frac{m_\mathscr{i}}{m_s}\hbar\big)$.

\section{Kinematics}
\begin{figure}[h!]
\vspace{-0.3cm}
    \centering
    \includegraphics[width=\columnwidth]{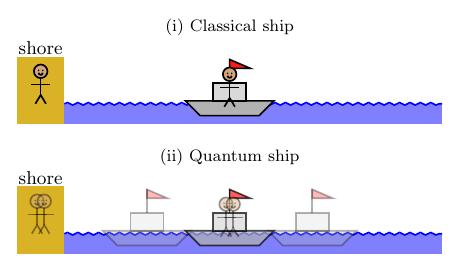}
    \caption{Classical and quantum ships illustrating the extension of the relativity principle.}
    \label{Fig2}
\end{figure}

To motivate the extension of the principle of relativity, consider a classical ship moving uniformly relative to the shore, as shown in Fig.~\ref{Fig2}. According to the relativity principle, an observer inside the ship obtains identical experimental results whether the ship is at rest or in uniform motion. This means that no experiment performed entirely within the ship can reveal its absolute velocity, and the dynamical laws are independent of this.

When the classical ship is replaced by a quantum ship, as in Fig.~\ref{Fig2}, its state of motion is no longer a c-number but a quantity that may be uncertain or entangled with other systems. Consequently, an observer transformation can no longer be regarded as a passive coordinate transformation. Moreover, since the physical predictions are associated with transition probability amplitudes, and the kinematic accessibility of the observer's proper state of motion is related to the algebraic structure of observables, then a relativity principle —if it exists— between these quantum observables must be extended to that structure level.

Accordingly, we introduce the \emph{quantum principle of relativity}  by imposing two fundamental requirements for \emph{general observers} - that is, quantum and classical observers. 
\begin{itemize}
\item[$\mathscr{R}_1$:] General observers are equivalent and must have the same physical predictions.
\item[$\mathscr{R}_2$:] A general observer cannot access its own quantum state of motion.  
\end{itemize}
These two conditions are sufficient to provide a consistent formulation of quantum theory for general observers, extending the conventional framework. This extension is conceptually important because it removes the implicit classical idealization underlying standard quantum mechanics.

Hereafter, the theory is formulated in the non-relativistic, one-dimensional continuum limit; detailed calculations and its three-dimensional extension appear in the Appendix.

\subsubsection{The General Observer Transformations} The principle
$\mathscr{R}_1$ implies that if a general observer $s$ describes the world in a state $\ket{\Psi}^s$ or $\ket{\Phi}^s$, then another general observer $s'$ will describe the world in states $\ket{\Psi'}^{s'}$ or $\ket{\Phi'}^{s'}$, respectively, under the requirement that the two are equivalent and produce the same physical predictions. Then,
\begin{equation}
\label{<phi|psi>=<phi'|psi'>equation}
    \braket{\Phi|\Psi}^s=\braket{\Phi'|\Psi'}^{s'}.
\end{equation}
This establishes an isometric map \footnote{See Wigner's Theorem for details on symmetry transformations between two Hilbert spaces, $\text{dim}(\mathscr{H}^s)=d_s$, $\text{dim}(\mathscr{H}^{s'})=d_{s'}$, in \cite{Bargmann:1964zj}.} connecting the observer-dependent Hilbert spaces as
$\mathscr{T}_s^{s'} : \mathscr{H}^s \rightarrow \mathscr{H}^{s'}$, with states and operators transforming according to
\begin{equation}
\label{O->O'equation}
\mathscr{T}_s^{s'} \ket{\Psi}^s=\ket{\Psi'}^{s'},\hspace{0.3cm}\mathscr{T}^{s'}_s\mathcal{O}^s\mathscr{T}^s_{s'}=\mathcal{O}^{s'},
\end{equation}
respectively. Note that each Hilbert space describes the corresponding external observers and particles, accommodating their quantum properties such as superposition or entanglement. Requiring that the position and momentum expectation values for the particle $\mathscr{i}$ transform according to the classical relations
\begin{equation*}
    \begin{split}
        &\braket{\hat{x}_\mathscr{i}^s}_{\Psi} = \braket{\hat{x}_\mathscr{i}^{s'}}_{\Psi'} - \braket{\hat{x}_s^{s'}}_{\Psi'},\\
        &\braket{\hat{p}_\mathscr{i}^s}_{\Psi} = \braket{\hat{p}_\mathscr{i}^{s'}}_{\Psi'} - \frac{m_\mathscr{i}}{m_s} \braket{\hat{p}_s^{s'}}_{\Psi'},\\
        &\braket{\hat{x}_{s'}^s}_{\Psi} =-\braket{\hat{x}_s^{s'}}_{\Psi'},\\
        &\braket{\hat{p}_{s'}^s}_{\Psi} =-\frac{m_{s'}}{m_s} \braket{\hat{p}_s^{s'}}_{\Psi'},        
    \end{split}
\end{equation*}
defines the general observer transformations $\mathscr{T}_s^{s'}$, which act on position and momentum operators as
\begin{equation}
\label{TXT'->X' and TPT'->P' equations}
\begin{aligned}
&\mathscr{T}_s^{s'} (\hat{x}_\mathscr{i}^s\otimes\hat{1}_{s'}^{s}) \mathscr{T}_{s'}^{s}=  \hat{x}_\mathscr{i}^{s'}\otimes\hat{1}_{s}^{s'}-\hat{1}_\mathscr{i}^{s'}\otimes\hat{x}_{s}^{s'},\\
&\mathscr{T}_s^{s'} (\hat{p}_\mathscr{i}^s\otimes\hat{1}_{s'}^s) \mathscr{T}_{s'}^{s}=  \hat{p}_\mathscr{i}^{s'}\otimes\hat{1}_{s}^{s'}-\frac{m_\mathscr{i}}{m_s}\hat{1}_\mathscr{i}^{s'}\otimes\hat{p}_{s}^{s'},\\
&\mathscr{T}_s^{s'} (\hat{1}_\mathscr{i}^s\otimes\hat{x}_{s'}^s) \mathscr{T}_{s'}^{s}=-\hat{1}_\mathscr{i}^{s'}\otimes\hat{x}^{s'}_s,\\
&\mathscr{T}_s^{s'} (\hat{1}_\mathscr{i}^s\otimes\hat{p}_{s'}^{s}) \mathscr{T}_{s'}^{s}= -\frac{m_{s'}}{m_s}
\hat{1}_\mathscr{i}^{s'}\otimes\hat{p}_{s}^{s'},    
\end{aligned}
\end{equation}
where, throughout this work, the notations for observables $\hat{\mathscr{o}}_i^{(obs)}\equiv\hat{\mathscr{o}}_i^{(obs)}\otimes \hat{1}_{(obs')}^{(obs)}$ and $\hat{\mathscr{o}}_{(obs')}^{(obs)}\equiv\hat{1}_i^{(obs)}\otimes \hat{\mathscr{o}}_{(obs')}^{(obs)}$ are used. These transformations extend the classical transformation to include superpositions of states.

For example, suppose the state of $s'$ as measured by $s$ is given by a coherent superposition of two distinct positions
\begin{equation}
\ket{\varphi}_{s'}^{s}=\frac{1}{\sqrt{2}}\ket{c}_{s'}^s+\frac{1}{\sqrt{2}}\ket{-c}_{s'}^s.
\end{equation}
Then, $\mathscr{T}_s^{\varphi}$ cannot be generated by a classical translation, as it corresponds to a superposition of distinct observer displacements.

\subsubsection{Quantization Rules for General Observers} 
The principle $\mathscr{R}_2$ implies precise constraints on physical operators. As described in \cite{Aharonov:1984zz}, consider the particle operators $\hat{x}^s_\mathscr{i}$ and $\hat{p}^s_\mathscr{i}$ of particle $\mathscr{i}$ (mass $m_\mathscr{i}$), measured by observer $s$ (mass $m_s$), and the center-of-mass operators $\hat{x}^{s'}_{\mathrm{cm}}$ and $\hat{p}^{s'}_{\mathrm{cm}}$ of the combined ‘‘particle ${\mathscr{i}}$- observer $s$’’ system, measured by an external observer $s'$ (mass $m_{s'}$). Therefore, since $s$ cannot tell whether it is in a position or momentum eigenstate with respect to $s'$, the above observables can be sharply measurable, leading to
\begin{equation}
\label{[x,v]=0equation}
    [\hat{x}^s_\mathscr{i},\hat{p}^{s'}_{\mathrm{cm}}]=[\hat{x}^{s'}_{\mathrm{cm}},\hat{p}^s_\mathscr{i}]=0.
\end{equation}
This condition is violated in standard quantum mechanics when $s$ and $s'$ are treated as quantum objects quantized by the canonical commutation rules. Hence, to enforce consistency with $\mathscr{R}_2$, the quantization prescription relative to a general observer $s$ of mass $m_s$ must be
\begin{equation}
\begin{split}
    \label{[x,p]general}
[\hat{x}_\mathscr{i}^s&, \hat{p}_\mathscr{j}^s] = i \hbar \left( \delta_{\mathscr{i}\mathscr{j}} + \frac{m_\mathscr{j}}{m_s} \right),\\
&[\hat{x}_\mathscr{i}^s, \hat{x}_\mathscr{j}^s]= [\hat{p}_\mathscr{i}^s, \hat{p}_\mathscr{j}^s]=0,
\end{split}
\end{equation}
where the indices $\mathscr{i}$ and $\mathscr{j}$ indicate the quantum objects of mass $m_\mathscr{i}$, and $m_\mathscr{j}$, respectively. 

These new quantization rules have the following physical interpretation: 
\begin{enumerate}[leftmargin=11pt,label=\roman*.]
    \item The additional term encodes the intrinsic quantum fluctuations of the observer. 
    
    \item They are compatible with general observer transformations defined in Eq.~\eqref{TXT'->X' and TPT'->P' equations}. Thus, applying the general observer transformation to standard quantization rules gives $\braket{\hat{p}_\mathscr{i}^s}_{\Psi} \neq \braket{\hat{p}_\mathscr{i}^{s'}}_{\Psi'} - \frac{m_\mathscr{i}}{m_s} \braket{\hat{p}_s^{s'}}_{\Psi'}$.
    
    \item They are relative on both sides, as opposed to $[\hat{x}_\mathscr{k}^s,\hat{p}_\mathscr{k}^s]=i\hbar$, which combines relative and absolute quantities. Although equations of the form $\mathscr{E}^2-\mathscr{p}^2=\mathscr{m}^2$ are similar in this respect to the canonical commutation, they still contain an intrinsic particle parameter like $\mathscr{m}$ instead of the absolute value $\hbar$.

    \item Their non-canonical nature prevents the separation of the quantum objects $\mathscr{i}$ and $\mathscr{j}$ into completely independent Hilbert spaces relative to $s$ when $m_s$ is finite. Specifically, the Hilbert space $\mathscr{H}^s$ is given by
    \begin{equation}
    \begin{split}
        &\mathscr{H}^s=\mathscr{H}^s_{x_\mathscr{i}}\otimes \mathscr{H}^s_{x_{s'}}=\mathscr{H}^s_{p_\mathscr{i}}\otimes \mathscr{H}^s_{p_{s'}}, \\
        &\mathscr{H}^s_{x_\mathscr{i}}\neq\mathscr{H}^s_{p_\mathscr{i}}, \mathscr{H}^s_{x_{s'}}\neq\mathscr{H}^s_{p_{s'}}.
    \end{split}
    \end{equation}
    Thus, operators written as $\hat{x}_\mathscr{i}^s\otimes\hat{1}_{s'}^{s}$ or $\hat{p}_\mathscr{i}^s\otimes\hat{1}_{s'}^{s}$ correspond to different factorizations of the total Hilbert space, $\mathscr{H}^s_{x_\mathscr{i}}\otimes \mathscr{H}^s_{x_{s'}}$, and $\mathscr{H}^s_{p_\mathscr{i}}\otimes \mathscr{H}^s_{p_{s'}}$, respectively.

   \item They reduce to the standard canonical rule in the \textit{classical-observers limit} where observer fluctuations are ignored:  when observers are infinitely massive and relatively localized, and/or when observers' relative positions and momenta are taken as c-numbers. 
\end{enumerate}

\subsubsection{Properties of the General Observer Transformations and Quantization Rules}
\begin{enumerate}[leftmargin=11pt,label=\roman*.]
\item For multiple observers, the general observer transformations obey the composition property 
\begin{equation}
    \mathscr{T}_{s_1}^{s_2} \mathscr{T}_{s_2}^{s_3}\ldots  \mathscr{T}_{s_{n-1}}^{s_n}= \mathscr{T}_{s_1}^{s_n}.
\end{equation}
This property allows the straightforward extension of results stated for two observers, $s$ and $s'$, to an arbitrary number of observers.

\item
The general observer transformation definition given in Eq.~\eqref{TXT'->X' and TPT'->P' equations} is equivalent to
\begin{equation}
\label{property2}
\begin{split}
\mathscr{T}_s^{s'}&\equiv\int \ket{x'_\mathscr{i}, x'_{s}  }^{s'}\bra{x_\mathscr{i},x_{s'}}^{s}
dx_\mathscr{i}dx_{s'},\\
\mathscr{T}_s^{s'} &\equiv\int \ket{ p'_\mathscr{i},  p'_{s}  }^{s'} \bra{p_\mathscr{i},p_{s'}   }^{s}
dp_\mathscr{i}dp_{s'},
\end{split}
\end{equation}
with
\begin{equation}
\label{quantumcoordinatetransformations}
\begin{aligned}
    &x'_\mathscr{i}= x_\mathscr{i}-x_{s'}, & x'_s&=-x_{s'},   \\
    &p'_\mathscr{i}= p_\mathscr{i}-{ \frac{ m_\mathscr{i}}{m_{s'}}} p_{s'}, &p'_s&=-{\frac{m_s}{m_{s'}}}p_{s'}.
\end{aligned}    
\end{equation}

\item The new quantization rules are covariant under  $\mathscr{T}_s^{s'}$. Explicitly, for a single particle $\mathscr{i}$, the commutator in $s'$ is given by
\begin{equation}
\label{property3part1}
[\hat{x}_\mathscr{i}^{s'}, \hat{p}_\mathscr{i}^{s'}] = i \hbar \left(1 + \frac{m_\mathscr{i}}{m_{s'}}\right),
\end{equation}
and for two distinct particles $\mathscr{i}\neq\mathscr{j}$, the commutator is
\begin{equation}
\label{property3part2}
[\hat{x}_\mathscr{i}^{s'}, \hat{p}_{\mathscr{j}}^{s'}] = i \hbar \frac{m_{\mathscr{j}}}{m_{s'}}.
\end{equation}

\item The momentum operators in the position representation are modified as
\begin{equation}
\label{newmomentums}
\begin{split}
\hat{p}_{\mathscr{i}}^s
    &=-i\hbar\!\left[\!
    \left(1+\frac{m_\mathscr{i}}{m_s}\right)\partial_{x_\mathscr{i}}
    +\frac{m_\mathscr{i}}{m_s}\partial_{x_{s'}}
    \right],\\
\hat{p}_{s'}^s
    &=-i\hbar\!\left[\!
    \frac{m_{s'}}{m_s}\partial_{x_\mathscr{i}}
    +\left(1+\frac{m_{s'}}{m_s}\right)\partial_{x_{s'}}
    \right],
\end{split}
\end{equation}
reflecting that they are not pure translations of one specific particle.

\item In the absence of $s'$, one can still define $\mathscr{T}_s^\mathscr{i}$ by treating symmetrically the particle $\mathscr{i}$ as an observer. In this case, the transformation reduces to
\begin{equation}
\label{defintionofTsi}
\begin{split}
\mathscr{T}_s^{\mathscr{i}}&\equiv\int \ket{x_s  }^{\mathscr{i}}\bra{x_\mathscr{i}}^{s}
dx_\mathscr{i}=\int \ket{ p_s }^{\mathscr{i}} \bra{p_\mathscr{i}}^{s}dp_\mathscr{i},\\
\end{split}
\end{equation}
with $x_s=-x_\mathscr{i}$ and $p_s=-\frac{m_s}{m_\mathscr{i}}p_\mathscr{i}$, and
\begin{equation}
\label{newmomentumssingle}
\begin{split}
       &\hat{p}_{\mathscr{i}}^s=-i\hbar\left(1+\frac{m_\mathscr{i}}{m_s}\right)\partial_{x_\mathscr{i}},\\   &\hat{p}^{\mathscr{i}}_s=-i\hbar\left(1+\frac{m_s}{m_\mathscr{i}}\right)\partial_{x_s}.
\end{split}
\end{equation}

\end{enumerate}

\subsubsection{An Example of a General Observer Transformation}
Consider the state $\ket{\Psi}^s$ and its image, $\ket{\Psi'}^{s'}=\mathscr{T}_s^{s'}\ket{\Psi}^s$, relative to observers $s$ and $s'$, respectively, that can be written as
\begin{equation}
\label{exampleoftransformation_positionspace}
\begin{aligned}
 &\ket{\Psi}^{s}&=&\int \psi^s_\mathscr{i}(x_\mathscr{i})\phi^s_{s'}(x_{s'})\ket{x_\mathscr{i},x_{s'}}^s dx_\mathscr{i}dx_{s'},\\
&\ket{\Psi'}^{s'}&=&\int \psi^{s'}_{\mathscr{i}}(x'_{\mathscr{i}}, x'_{s})\phi^{s'}_{s}(x'_{s})\ket{x_\mathscr{i}',x'_s}^{s'}dx'_\mathscr{i} dx'_s,     
\end{aligned}
\end{equation}
with the ‘‘natural’’ wave functions relative to $s'$ defined as $\psi^{s'}_{\mathscr{i}}(x'_{\mathscr{i}}, x'_{s})=\psi^s_{\mathscr{i}}(x'_\mathscr{i}-x'_{s})$, $\phi^{s'}_{s}(x'_{s})=\phi^s_{s'}(-x'_{s})$. In the momentum basis, these states are
\begin{equation}
\label{exampleoftransformation_momentumspace}
\begin{aligned}
    &\ket{\Psi}^{s}&=&\int \tilde{\psi}_\mathscr{i}^s(\pi_\mathscr{i})  \tilde{\phi}_{s'}^s(\pi_{s'}) \ket{p_\mathscr{i},p_{s'} }^sdp_\mathscr{i}dp_{s'},\\
    &\ket{\Psi'}^{s'}&=&\int \tilde{\psi}_\mathscr{i}^{s}(\pi'_\mathscr{i})  \tilde{\phi}^{s'}_s(\pi'_{s}+\pi'_\mathscr{i}) \ket{p'_\mathscr{i},p'_{s} }^{s'}dp'_\mathscr{i}dp'_{s},
\end{aligned}
\end{equation}
with $\tilde{}$ denoting the standard Fourier transformation, the auxiliary momenta defined by $\pi_\mathscr{k}=p_\mathscr{k}-\frac{m_\mathscr{k}\sum_\mathscr{l} p_\mathscr{l}}{m_\mathscr{total}}$ ($m_\mathscr{total}$ as the total mass of the system including the observer) and the coordinate transformations of Eq.~\eqref{quantumcoordinatetransformations}. 
From this example, we enumerate some key properties:

\begin{enumerate}[leftmargin=11pt,label=\roman*.]

\item Entanglement is a frame-dependent property, as we can see in Eq.~\eqref{exampleoftransformation_positionspace}, for $s'$, the wavefunction related to $\mathscr{i}$ will depend on $x'_s$.

\item The unentangled state in the position basis, $\Psi^s(x_\mathscr{i},x_{s'})=\psi^s_\mathscr{i}(x_\mathscr{i})\phi^s_{s'}(x_{s'})$ will generally be entangled in the momentum basis even for the same reference system $s$, $\Psi^s(p_\mathscr{i},p_{s'})=\tilde\psi^s_\mathscr{i}(\pi_\mathscr{i})\tilde\phi^s_{s'}(\pi_{s'})$. This follows from the non-canonical nature of the quantization rules in Eq.~\eqref{[x,p]general}.

\item For the case of the localized observer $s'$ with respect to $s$, as $\phi_{s'}^s(-x'_s)=\delta(x'_s-c)$, it results in
\begin{equation}
    \begin{aligned}
     &\ket{\Psi}^{s}&=&\int   {\psi}_\mathscr{i}^s(x_\mathscr{i})   \delta(x_{s'}-c)  \ket{x_\mathscr{i},x_{s'}}^s dx_\mathscr{i}dx_{s'},\\
    &\ket{\Psi'}^{s'}&=&\int   {\psi'}_\mathscr{i}^{s'}(x'_\mathscr{i})   \delta(x'_s-c')  \ket{x'_\mathscr{i},x'_{s}}^{s'} dx'_\mathscr{i}dx'_{s},\\    
    &\ket{\Psi}^{s}&=&\int \tilde{\psi}_\mathscr{i}^s(\pi_\mathscr{i})  e^{i\pi_{s'}c/\hbar} \ket{p_\mathscr{i},p_{s'} }^{s}dp_\mathscr{i}dp_{s'},\\
    &\ket{\Psi'}^{s'}&=&\int \tilde{\psi'}_\mathscr{i}^{s'}(\pi'_\mathscr{i})  e^{i\pi'_{s}c'/\hbar} \ket{p'_\mathscr{i},p'_{s} }^{s'}dp'_\mathscr{i}dp'_{s}.     
    \end{aligned}
\end{equation}
Thus, after the general observer transformation $s\rightarrow s'$, a pure translation 
\begin{equation}
 {\psi'}_\mathscr{i}^{s'}(x'_\mathscr{i})={\psi}_\mathscr{i}^s(x'_\mathscr{i}+c'),   
\end{equation}
with $c'=-c$, and a superposition of boosts is implemented in the basis of position and momentum, respectively.

\item In the infinite mass limit, $m_{s},m_{s'}\rightarrow \infty$ with $m_{s}/{m_{s'}}$ finite, the momentum transforms as $p_\mathscr{i}'=p_\mathscr{i}+u_\mathscr{i}$ with $u_\mathscr{i}$ undetermined, and $\pi_\mathscr{i}$ reduces to $p_\mathscr{i}$, separating the relation between boosts and translations.

However, in this scenario, we can still have quantum properties if the observers are in a superposition of position states with respect to each other as 
\begin{equation}
    \ket{\phi}^s_{s'}=\int \phi(x_{s'})\ket{x_{s'}}^sdx_{s'}, \hspace{0.2cm}\phi(x_{s'})\neq\delta (x_{s'}-c).
\end{equation}
Then, localizing the observer $s'$ with respect to $s$ by imposing $\phi(x_{s'})=\delta (x_{s'}-c)$, the states relative to $s$ and $s'$ in the position and momentum representation, respectively, are
\begin{equation}
\begin{aligned}
     &\ket{\Psi}^{s}&=&\int   {\psi}_\mathscr{i}^s(x_\mathscr{i})   \delta(x_{s'}-c)  \ket{x_\mathscr{i},x_{s'}}^s dx_\mathscr{i}dx_{s'},\\
    &\ket{\Psi'}^{s'}&=&\int   {\psi'}_\mathscr{i}^{s'}(x'_\mathscr{i})   \delta(x'_s-c')  \ket{x'_\mathscr{i},x'_{s}}^{s'} dx'_\mathscr{i}dx'_{s},\\
    &\ket{\Psi}^{s}&=&\int \tilde{\psi}_\mathscr{i}^s(p_\mathscr{i})      e^{-\frac{ip_{\mathscr{i}}c}{2\hbar}} e^{\frac{ip_{s'}c}{2\hbar}}\ket{p_\mathscr{i},p_{s'} }^{s}dp_\mathscr{i}dp_{s'},\\
    &\ket{\Psi'}^{s'}&=&\int \tilde{\psi'}_\mathscr{i}^{s'}(p'_\mathscr{i})  e^{-\frac{ip_{\mathscr{i}}c'}{2\hbar}} e^{\frac{ip'_{s}c'}{2\hbar}} \ket{p'_\mathscr{i},p'_{s} }^{s'}dp'_\mathscr{i}dp'_{s},
\end{aligned}
\end{equation}
with $c'=-c$.

This corresponds to the classical-observers limit, in which translations and boosts become independent, and the standard unitary transformations relating the $s$ and $s'$ descriptions are recovered, with the states and observables transforming as
\begin{equation}
    \ket{\psi'}^{s'}=U\ket{\psi}^s, \quad\mathcal{O'}^{s'}=U\mathcal{O}^{s}U^\dagger.
\end{equation}

\item If a projection in the particle $\mathscr{i}$ is ‘‘performed’’ by $s$, then it induces a projection of the entire system relative to $s'$. To illustrate this, consider a projection of $\mathscr{i}$ in the position basis of $s$ given by
\begin{equation}
\label{projection1}
\begin{split}
    &\hat {{\text{P}}}_\mathscr{i}^s=\ket{x_a}\bra{x_a}_{\mathscr{i}}^s\otimes \hat1_{s'}^s,\\
    &\hat {{\text{P}}}_\mathscr{i}^s\ket{\Psi}^s=\int \psi_\mathscr{i}^s(x_a)\phi_{s'}^s(x_{s'})\ket{x_a,x_{s'}}^sdx_{s'}.
\end{split}
\end{equation}
From the perspective of $s'$, it turns into another projection for the whole $s-\mathscr{i}$ system  as
\begin{equation}
\label{projection2}
\begin{split}
       &{\hat {{\text{P}}}}_{\mathscr{i},s}^{s'}=\mathscr{T}^{s'}_s \hat {{\text{P}}}_\mathscr{i}^s(x_a)  \mathscr{T}^s_{s'}\\
       &=\int\ket{x'_a,x'_s}\bra{x'_a,x'_s}^{s'}dx'_s,\hspace{0.15cm}\text{with }x'_a=x'_s+x_a.\\
 \end{split}
\end{equation}

In the classical-observers limit, since $\mathscr{T}_s^{s'}$ reduces to the standard unitary transformations on the particle $\mathscr{i}$, then it is possible to write:
\begin{equation}
\label{projection3}
    {\hat {{\text{P}}}}_{\mathscr{i},s}^{s'}\rightarrow     {\hat {{\text{P}}}}_{\mathscr{i}}^{s'}=\ket{x'_a}\bra{x'_a}_{\mathscr{i}}^{s'}\otimes \hat1_{s}^{s'},
\end{equation}
indicating the projection is universal for all observers.
\end{enumerate}

\newpage

\section{Dynamics}
The principle of relativity, $\mathscr{R}_1$, requires that the Hamiltonian transforms according to
\begin{equation}
\label{Hresctrited by R2}
    \mathcal{H}^{s'}=\mathscr{T}^{s'}_s \mathcal{H}^{s} \mathscr{T}^{s}_{s'}, \hspace{0.5cm}\forall s'(\textit{general observers}).
\end{equation}
Hence, the evolution of a state $\ket{\Psi}^s$ relative to the observer $s$ is given by
\begin{equation}
\label{Hamiltonian}
\mathcal{H}^s\ket{\Psi}^s=i\hbar\frac{\partial}{\partial t}\ket{{\Psi}^s}.
\end{equation}

\subsubsection{Examples of Hamiltonians}
For an arbitrary number of general observers, Hamiltonians consistent with the observer symmetry of Eq.~\eqref{Hresctrited by R2} include:
\begin{enumerate}[leftmargin=11pt,label=\roman*.]
    \item The $N$ free-particle case,
\begin{equation}
\label{Case1FreeHamiltnoian}
[\mathcal{H}^s_{\mathcal{free}}]_N = \sum_{\mathscr{i}=1}^{N} \frac{[\hat p^s_\mathscr{i}]^2}{2 m_\mathscr{i}} + \sum_{s'}\frac{[\hat p_{s'}^s]^2}{2 m_{s'}} 
- \frac{[\hat P_{\mathcal{total}}^s]^2}{2 M_{\mathcal{total}}}, 
\end{equation}
with $\hat P_{\mathcal{total}}^s= \sum_{s'}\hat p_{s'}^s +\sum_{{\mathscr{i}}=1}^{N} \hat p_\mathscr{i}^s  $, and $M_{\mathcal{total}}=m_s + \sum_{s'} m_{s'} + \sum_{\mathscr{i}=1}^{N} m_\mathscr{i}$. The last term of Eq.~\eqref{Case1FreeHamiltnoian} removes the center of mass momentum of the whole system, ensuring Eq.~\eqref{Hresctrited by R2} is satisfied. The dynamics of expectation values are given by
\begin{equation}
\label{Ehrenfest1}
   \frac{d \braket{\hat x^s_\mathscr{k}}}{dt}=\frac{ \braket{\hat p^s_\mathscr{k}}}{m_\mathscr{k}},\hspace{0.5cm}\frac{d \braket{\hat p^s_\mathscr{k}}}{dt}=0,\hspace{0.5cm}\forall \mathscr{k}.
\end{equation}

\item The two-particle interaction case,
\begin{equation}
\label{Case2InteractionHamiltnoian}
\mathcal{H}^s_2 =[\mathcal{H}^s_{\mathcal{free}}]_2 +\hat V({\hat x_\mathscr{a}}-{\hat x_\mathscr{b}}),
\end{equation} 
with the dynamics of expectation values given by
\begin{equation}
\label{Ehrenfest2}
    \begin{aligned}
    &\frac{d \braket{\hat x^s_\mathscr{k}}}{dt}=\frac{ \braket{\hat p^s_\mathscr{k}}}{m_\mathscr{k}},&\hspace{0.3cm}&\frac{d \braket{\hat p^s_\mathscr{k}}}{dt}=0,\hspace{0.3cm}\forall \mathscr{k}\neq\mathscr{a},\mathscr{b};\\
    &\frac{d \braket{\hat x^s_\mathscr{a}}}{dt}=\frac{ \braket{\hat p^s_\mathscr{a}}}{m_\mathscr{a}},&\hspace{0.3cm}&\frac{d \braket{\hat p^s_\mathscr{a}}}{dt}=-\frac{ \braket{\partial^s_\mathscr{a}\hat V}}{m_\mathscr{a}},\\       
    &\frac{d \braket{\hat x^s_\mathscr{b}}}{dt}=\frac{ \braket{\hat p^s_\mathscr{b}}}{m_\mathscr{b}},&\hspace{0.3cm}&\frac{d \braket{\hat p^s_\mathscr{b}}}{dt}=-\frac{ \braket{\partial^s_\mathscr{b}\hat V}}{m_\mathscr{b}}.
    \end{aligned}
\end{equation}
\end{enumerate}

Note that satisfying Eq.~\eqref{Hresctrited by R2} is necessary to reproduce classical-like dynamics for expectation values via the Ehrenfest theorem as Eqs.~\eqref{Ehrenfest1} and \eqref{Ehrenfest2}. For example, the standard free-particle Hamiltonian of $\mathscr{i}$ which is not symmetric under $\mathscr{T}^{s}_{\mathscr{i}}$, 
\begin{equation*}
    \mathcal{H}^s= \frac{\hat p_\mathscr{i}^2}{2  m_\mathscr{i}}, \hspace{0.35cm}\frac{d}{dt}\mathcal{O}^s=\frac{1}{i\hbar}[\mathcal{O}^s,\mathcal{H}^s],
\end{equation*}
leads to inconsistent expectation value dynamics using the momentum defined in Eq.~\eqref{newmomentumssingle}
\begin{equation}
     \frac{d}{dt}\braket{\hat {{x}}^s_\mathscr{i}}=\frac{\braket{\hat{p}_\mathscr{i}^s}}{m_\mathscr{i}}\bigg(1+\frac{m_\mathscr{i}}{m_s}\bigg).
\end{equation}

\subsubsection{Generalized Galilean Symmetry} The observer transformations $\mathscr{T}_{s}^{s'}$ that preserve the dynamics extend the conventional symmetry for classical inertial observers to general inertial observers. They satisfy
\begin{equation}
  \mathscr{G}_s^{s'}:  [\mathcal{H}^{s},\hat{p}^s_\mathscr{k}]=[\mathcal{H}^{s'},\hat{p}^{s'}_\mathscr{k}], \hspace{0.5cm}\forall \mathscr{k},
\end{equation}
or explicitly, $\mathscr{T}_s^{s'}[\mathcal{H}^{s'},\hat{p}^{s'}_\mathscr{k}]\mathscr{T}_s^{s'}=[\mathcal{H}^{s},\hat{p}^{s}_\mathscr{k}-\frac{m_\mathscr{k}}{m_s}\hat{p}^s_{s'}]$, which holds if
\begin{equation}
    [\mathcal{H}^s,\hat{p}^s_{s'}]=0 \rightarrow \mathscr{T}_s^{s'}\equiv\mathscr{G}_s^{s'}.
\end{equation}
From the two previous examples, the free-particle Hamiltonian respects $\mathscr{G}_{s_\mathscr{i}}^{s_\mathscr{j}}$; however, the interacting particles in Eq.~\eqref{Case2InteractionHamiltnoian} break the symmetry, since
\begin{equation}
    [\mathcal{H}^s,\hat{p}^s_{\mathscr{a,b}}]\neq0.
\end{equation}

\subsubsection{The Free-Particle Description} Consider a particle of mass $m_\mathscr{i}$ described by:
\begin{enumerate}[leftmargin=11pt,label=\roman*.]

\item Two general observers $s$ and $s'$, with the Hamiltonian 
\begin{equation}
    \mathcal{H}^s_{\mathscr{i}s'} = \frac{\hat p_\mathscr{i}^2}{2 m_\mathscr{i}} + \frac{\hat p_{s'}^2}{2 m_{s'}} 
- \frac{(\hat p_\mathscr{i}+\hat p_{s'})^2}{2(m_s +m_{s'} +m_\mathscr{i})},
\end{equation}
that can be solved by $\upvarphi(x_\mathscr{i},x_{s'},t) =e^{-\frac{iEt}{\hbar}}\varphi(x_\mathscr{i},x_{s'})$ satisfying 
\begin{equation}
-\bigg(\frac{\hbar^2}{2\mu_\mathscr{i}}\partial^2_{x_{\mathscr{i}}}+
\frac{\hbar^2}{2\mu_{s'}}\partial^2_{x_{s'}}+
\frac{\hbar^2}{m_s}\partial_{x_{\mathscr{i}}}\partial_{x_{s'}}\bigg)\varphi
=E \varphi,
\end{equation}
where the reduced masses are defined relative to the observer $s$, as $ 1/\mu_\mathscr{k} = 1/m_\mathscr{k}+1/m_s$. Then, the general solution in the position basis is 
\begin{equation}
\upvarphi(x_\mathscr{i},x_{s'},t)= \int F(\tilde p_\mathscr{i},\tilde p_{s'})e^{-\frac{iEt}{\hbar}}e^{\frac{i\tilde p_\mathscr{i} x_\mathscr{i}}{\hbar}}e^{\frac{i\tilde p_{s'} x_{s'}}{\hbar} }d\tilde{p}_\mathscr{i}d\tilde{p}_{s'},
\end{equation}
with $E=\frac{\tilde{p}_\mathscr{i}^2}{2\mu_\mathscr{i}}+
\frac{\tilde{p}_{s'}^2}{2\mu_{s'}}+
\frac{\tilde{p}_\mathscr{i}\tilde{p}_{s'}}{m_s}$.

\item One general observer $s$, with the Hamiltonian
\begin{equation}
        \mathcal{H}^s=  \frac{\hat p_\mathscr{i}^2}{2 \tilde m_\mathscr{i}},\hspace{0.5cm}\tilde m_\mathscr{i}=m_\mathscr{i}\bigg(1+\frac{m_\mathscr{i}}{m_s}\bigg).
\end{equation} 
Using $\upvarphi_i^s(x_\mathscr{i},t)=e^{-\frac{iEt}{\hbar}}\varphi_i^s(x_\mathscr{i})$ and Eq.~\eqref{newmomentumssingle}, results in
\begin{equation}
\label{1freeparticleschrodinger}
     -\frac{\hbar^2}{2 \mu}\partial^2_{x_{\mathscr{i}}}\varphi(x_{\mathscr{i}})=E \varphi(x_{\mathscr{i}}), \hspace{0.5cm}\frac{1}{\mu}=\frac{1}{m_\mathscr{i}}+\frac{1}{m_s}.
\end{equation}
Then, for the observer $s$, it corresponds to the standard dynamics of a single free particle with reduced mass $\mu$. However, if the particle $\mathscr{i}$ is taken as the observer, it gets the same description, as the reduced mass remains identical. This reflects the existence of a Galilean symmetry between the observer $s$ and the particle $\mathscr{i}$, $\mathscr{G}_s^{\mathscr{i}}$.

\end{enumerate}

\section{Discussion}

\subsubsection{Implications on Interpretations of The Quantum Theory} 

In this framework, quantum states are always defined relative to an observer. Hence, a \textit{total} wavefunction that simultaneously describes both the observer and particle lacks a clear physical interpretation, apparently avoiding the \textit{measurement problem} of quantum mechanics, similarly to the relational interpretation of \cite{Rovelli:1995fv}. 

Nevertheless, notions such as measurement, apparatus, and collapse remain conceptually ambiguous, and therefore, the problem is not fully resolved. A detailed clarification will be presented in a forthcoming work.

\subsubsection{Implications on the Wigner's Friend Paradox}

An apparent tension of the standard quantum mechanics arises in the ‘‘Wigner's friend’’ thought experiment \cite{Wigner1995}. Consider Wigner's friend ($F$) performs a projective measurement on a particle $\mathscr{i}$, obtaining a definite outcome. At the same time, Wigner ($W$), who has not measured anything, assigns a superposition of different outcome states to the joint system $F+\mathscr{i}$. How can $F$ get a definite outcome while $W$ simultaneously sees a superposition? 

Here, the relativity principle $\mathscr{R}_1$ restricts the interpretation of events. In a minimal setup of the scenario described above, the states described for $F$ are
\begin{equation}
    \begin{split}
    \textit{before: }    &\ket{\Psi}^F=\big(\alpha \ket{0}_\mathscr{i}^F+\beta \ket{1}_\mathscr{i}^F \big)\ket{0}^F_W,\\
    \textit{after: }    &\ket{\Phi}^F=\ket{0}_\mathscr{i}^F\ket{0}^F_W,\\
    \end{split}
\end{equation}
while for $W$,
\begin{equation}
    \begin{split}
    \textit{before: }    &\ket{\Psi'}^W=\big(\alpha \ket{0}_\mathscr{i}^W+\beta \ket{1}_\mathscr{i}^W \big)\ket{0}^W_F,\\
    \textit{after: }    &\ket{\Phi'}^W=\alpha \ket{0}_\mathscr{i}^W\ket{0}_F^W+\beta \ket{1}_\mathscr{i}^W\ket{1}_F^W,\\
    \end{split}
\end{equation}
thus, violating the relativity requirement for $\alpha\neq0,1$ as
\begin{equation*}
    \begin{split}
        \braket{\Phi|\Psi}^F&\neq\braket{\Phi'|\Psi'}^W,\\
        \alpha&\neq|\alpha|^2.
    \end{split}
\end{equation*}

Assuming a projection is ‘‘performed’’ by $F$, the present framework removes this tension by giving a consistent interpretation of events in two limiting cases:
\begin{enumerate}[leftmargin=11pt,label=\roman*.]
\item \textit{Classical observers.-} In this limit, $\mathscr{R}_1$ enforces the collapse in all classical frames, as shown in Eq.~\eqref{projection3}. Since no quantum properties are associated with observers, the states before and after the measurement for $F$ and $W$ are
    \begin{equation}
        \begin{aligned}
        &\ket{\Psi}^F=\alpha \ket{0}_\mathscr{i}^F+\beta \ket{1}_\mathscr{i}^F ,&   &\ket{\Phi}^F=\ket{0}_\mathscr{i}^F,\\
        &\ket{\Psi'}^W=\alpha \ket{0'}_\mathscr{i}^W+\beta \ket{1'}_\mathscr{i}^W,&   &\ket{\Phi'}^W= \ket{0'}_\mathscr{i}^W.\\
        \end{aligned}
\end{equation}

\item \textit{Quantum observers.-} 
In this limit, $\mathscr{R}_1$ constraints the state description of $F$ and $\mathscr{i}$, by enforcing the transition amplitudes. The collapse of $\mathscr{i}$ in $F$ corresponds to a collapse of $F$ and $\mathscr{i}$ in the $W$ description, as shown in Eqs.~\eqref{projection1} and \eqref{projection2}.

Additionally, if $W$ and $\mathscr{i}$ are unentangled for $F$, then $\mathscr{i}$ and $F$ are entangled for $W$, as illustrated in Eq.~\eqref{exampleoftransformation_positionspace}. Therefore, a minimal toy model capturing these features is given by
\begin{equation}
    \begin{split}
        &\ket{\Psi'}^W=\alpha' \ket{0'}_\mathscr{i}^W\ket{0'}^W_F+\beta' \ket{1'}_\mathscr{i}^W \ket{1'}^W_F,\\
        &\ket{\Phi'}^W=\alpha'' \ket{0'}_\mathscr{i}^W\ket{0'}^W_F+\beta'' \ket{1'}_\mathscr{i}^W \ket{1'}^W_F,\\
    \end{split}
\end{equation}
where the observer transformation $\mathscr{T}_F^W$ must ensure $\mathscr{R}_2$ with
\begin{equation}
\begin{split}
        \alpha=\alpha'\alpha''+\beta'\beta'',
\end{split}
\end{equation}
\end{enumerate}
In summary, the Wigner's friend paradox arises in the standard quantum mechanics because of the absence of a relativity rule between quantum observers; then, based solely on quantum mechanics, different observers could, in principle, assign ambiguous descriptions, such as projections or unitary evolutions,  that ultimately correspond to nonequivalent transition amplitudes, creating contrived interpretations.  
 
 By contrast, the quantum relativity principle enforces equality of transition amplitudes under the transformations $\mathscr{T}_s^{s'}$ between Wigner and his friend. In this context, a projection in one frame uniquely enforces the relative collapse for all observers in a schematic way without any additional interpretational assumptions.

\subsubsection{Implications on Uncertainty Relations} The new quantization rules imply modifications of the conventional uncertainty relations. 
Following the standard arguments, for a state $\ket{\Psi}^s$ relative to a general observer $s$ of mass $m_s$, the uncertainty in the position and momentum of a particle ${\mathscr{i}}$ of mass $m_{\mathscr{i}}$ satisfies
\begin{equation}
\label{Uncertainty1}
\Delta x_\mathscr{i}^s \, \Delta p_\mathscr{i}^s \ge \frac{\hbar}{2} \left( 1 + \frac{m_\mathscr{i}}{m_s} \right),
\end{equation}
while for two distinct particles $\mathscr{i}\neq\mathscr{j}$, a new uncertainty relation is given by
\begin{equation}
\label{Uncertainty2}
\Delta x_\mathscr{i}^s \, \Delta p_{\mathscr{j}}^s \ge \frac{\hbar}{2} \frac{m_{\mathscr{j}}}{m_s}.
\end{equation}
The covariance property of the quantization rules extends directly to the corresponding uncertainty relations. 

These novel uncertainties, which reduce to the standard form in the limit of an infinitely massive observer, reflect how the intrinsic fluctuations of the observer correlate different particles with possible consequences for precision measurements, mesoscopic systems, and quantum technologies.

\subsubsection{Implications on Experimental Signatures}
\begin{figure}[h!]
\vspace{-0.6cm}
    \centering
    \includegraphics[width=\columnwidth]{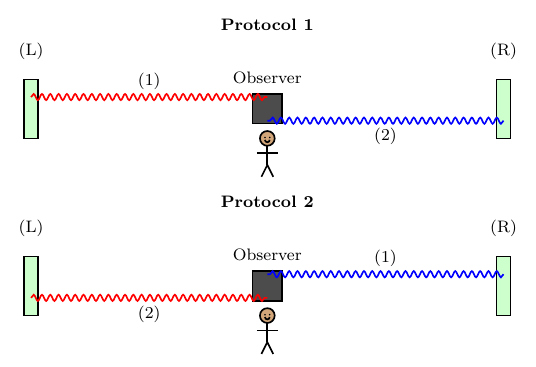}
    \caption{Illustrative representation of the two measurement protocols that an observer performs on two identical objects located to the left ($L$) and right ($R$). In protocol 1, $\hat{x}_L$ is measured before $\hat{p}_R$. In protocol 2, the order is reversed, with $\hat{p}_R$ measured before $\hat{x}_L$.}
    \label{Fig3}
\end{figure}

To test finite-mass corrections within this framework, consider an observer of mass $m_{\mathscr{O}}$ measuring two identical objects (e.g., mirrors) of mass $m_{\mathscr{M}}$ located to the left ($L$) and right ($R$), as shown in Fig.~\ref{Fig3}. Define two measurement protocols: in protocol 1, the observer first measures the position of $L$ ($\hat x_L$), then the momentum of $R$ ($\hat p_R$); in protocol 2, the order is reversed. The observer first measures the momentum of $R$ ($\hat p_R$), then the position of $L$ ($\hat x_L$).  In standard quantum mechanics, where the observer is effectively classical ($m_\mathscr{O} \to \infty$), the two protocols are operationally indistinguishable. 

In the present framework, however, Eq.~\eqref{Uncertainty2} implies that the two protocols are not equivalent, since $\hat{x}_L \hat{p}_R\neq\hat{p}_R \hat{x}_L$. This difference between the protocols leads to distinct joint probability distributions of these observables, which can be quantified by the covariance operator defined as
\begin{equation}
    \Delta C=\braket{\Delta \hat x_L\Delta \hat p_R}_2-\braket{\Delta \hat p_R\Delta \hat x_L}_1,
\end{equation}
where the subscripts label the protocol, and $\Delta \hat x_L=\hat{x}_L -\braket{\hat x_L}$, $\Delta \hat p_R=\hat{p}_R -\braket{\hat p_R}$. From Eq.~\eqref{Uncertainty2}, one finds
\begin{equation}
    \Delta C=i\hbar \frac{m_{\mathscr{M}}}{m_\mathscr{O}}.
\end{equation}
For concreteness, consider a large-scale interferometer setup in which $m_{\mathscr{M}}\sim 10\,\mathrm{kg}$, and $m_{\mathscr{O}}\sim 10^5$--$10^7\,\mathrm{kg}$ correspond approximately to the masses of the mirrors and the supporting structure, respectively. Then
\begin{equation}
    \left|\frac{\Delta C}{i\hbar}\right|= \frac{m_{\mathscr{M}}}{m_{\mathscr{O}}}\sim 10^{-4} - 10^{-6}.
\end{equation}

\subsubsection{Summary and
Outlook}
In the standard formulation of quantum mechanics, the canonical commutation relations are observer mass independent. This feature is consistent with the classical relativity principle, as one assumes the observers (or apparatus) are classical, so that they are described by c-numbers since their quantum fluctuations are neglected. 

Motivated by the need to describe physics in a finite universe, the present framework removes this classical-observer idealization. Observers are finite-mass quantum systems, and the relativity principle is extended to the level of transition amplitudes and the algebraic structure for observables- an extension we refer to as the quantum relativity principle. Its validity is not guaranteed, as one may argue that relativity applies only to classical observers; in this sense, it constitutes a new ingredient of the theory.

By imposing the quantum principle of relativity, we obtain a fully relative formulation of the quantum theory, in which the Hilbert spaces are observer-dependent and the relative quantization rules depend on the mass ratio between the particle and the observer. In this framework, it is therefore natural to conclude that quantization itself is a relative procedure, since the quantum fluctuations always contain this mass ratio. The canonical quantization emerges only in a classical-observers limit, where the mass of the observer is infinite. Nevertheless, we must emphasize that this new algebra can be defined in terms of certain canonical variables; however, these required canonical variables necessarily mix different system degrees of freedom and therefore do not represent the observables of each subsystem relative to a finite-mass observer.  

Additionally, the quantum relativity principle sheds light on interpretational problems that arise when observers both perform measurements and are themselves subject to measurement by other observers. Here, it is possible to provide a consistent interpretation of related paradoxes, such as the simple Wigner's friend scenario. Importantly, the finite-mass corrections implied by this formulation provide concrete setups in which these effects may become operationally relevant, as demonstrated by the proposed experimental test. Such a test would probe the observer's quantum fluctuations. 

The present framework can be regarded as a non-relativistic model of relative quantum kinematics without any external classical observers. Extending this construction to the relativistic regime, and ultimately to a cosmological quantum model incorporating gravity is an important direction for future work.

\section{Acknowledgments}
I am truly grateful to Manthos Karydas and Oscar Sumari for their insightful and stimulating discussions.

\bibliographystyle{unsrt}%Used BibTeX style is unsrt
\bibliography{mibibliografia}
\clearpage

\appendix

\section{Definitions of the general observer transformation, $\mathscr{T}_s^{s'}$}
Here, we prove the equivalence of definitions for the general observer transformation
\begin{equation}
\begin{split}
&\mathscr{T}_s^{s'} (\hat{x}_\mathscr{i}^s\otimes\hat{1}_{s'}^{s}) \mathscr{T}_{s'}^{s}=  \hat{x}_\mathscr{i}^{s'}\otimes\hat{1}_{s}^{s'}-\hat{1}_\mathscr{i}^{s'}\otimes\hat{x}_{s}^{s'},\\
&\mathscr{T}_s^{s'} (\hat{p}_\mathscr{i}^s\otimes\hat{1}_{s'}^s) \mathscr{T}_{s'}^{s}=  \hat{p}_\mathscr{i}^{s'}\otimes\hat{1}_{s}^{s'}-\frac{m_\mathscr{i}}{m_s}\hat{1}_\mathscr{i}^{s'}\otimes\hat{p}_{s}^{s'},\\
&\mathscr{T}_s^{s'} (\hat{1}_\mathscr{i}^s\otimes\hat{x}_{s'}^s) \mathscr{T}_{s'}^{s}=-\hat{1}_\mathscr{i}^{s'}\otimes\hat{x}^{s'}_s,\\
&\mathscr{T}_s^{s'} (\hat{1}_\mathscr{i}^s\otimes\hat{p}_{s'}^{s}) \mathscr{T}_{s'}^{s}= -\frac{m_{s'}}{m_s}
\hat{1}_\mathscr{i}^{s'}\otimes\hat{p}_{s}^{s'},    
\end{split}
\end{equation}
Then, by taking the left-hand side of the equations, we have for position
\begin{equation}
\begin{split}
    \mathscr{T}_s^{s'}& (\hat{x}_\mathscr{i}^s\otimes\hat{1}_{s'}^{s}) \mathscr{T}_{s'}^{s},\\
    &=\int x_\mathscr{i}\ket{x'_\mathscr{i},x'_{s}}\bra{x'_\mathscr{i},x'_{s}}^{s'}dx'_\mathscr{i} dx'_{s},\\
    &=\int (x'_\mathscr{i}-x'_{s})\ket{x'_\mathscr{i},x'_{s}}\bra{x'_\mathscr{i},x'_{s}}^{s'}dx'_\mathscr{i} dx'_{s},\\
    &=\int x'_\mathscr{i}\ket{x'_\mathscr{i}}\bra{x'_\mathscr{i}}^{s'}dx'_\mathscr{i}-\int x'_{s}\ket{x'_{s}}\bra{x'_{s}}^{s'}dx'_{s},\\
    &=\hat{x}_\mathscr{i}^{s'}\otimes\hat{1}_{s}^{s'}-\hat{1}_\mathscr{i}^{s'}\otimes\hat{x}_{s}^{s'}.\\
\end{split}
\end{equation}
Similarly, for the momentum operators, we get
\begin{equation}
\begin{split}
    \mathscr{T}_s^{s'}& (\hat{p}_\mathscr{i}^s\otimes\hat{1}_{s'}^{s}) \mathscr{T}_{s'}^{s},\\
    &=\int p_\mathscr{i}\ket{p'_\mathscr{i},p'_{s}}\bra{p'_\mathscr{i},p'_{s}}^{s'}dp'_\mathscr{i} dp'_{s},\\
    &=\int \bigg(p'_\mathscr{i}-\frac{m_\mathscr{i}}{m_s}p'_{s}\bigg)\ket{p'_\mathscr{i},p'_{s}}\bra{p'_\mathscr{i},p'_{s}}^{s'}dp'_\mathscr{i} dp'_{s},\\
    &=\int p'_\mathscr{i}\ket{p'_\mathscr{i}}\bra{p'_\mathscr{i}}^{s'}dp'_\mathscr{i}-\frac{m_\mathscr{i}}{m_s}
    \int p'_{s}\ket{p'_{s'}}\bra{p'_{s}}^{s'}dp'_{s},\\
    &=\hat{p}_\mathscr{i}^{s'}\otimes\hat{1}_{s}^{s'}-\frac{m_\mathscr{i}}{m_s}\hat{1}_{\mathscr{i}}^{s'}\otimes\hat{p}_{s}^{s'}.
\end{split}
\end{equation}
Moreover,
\begin{equation}
\begin{split}
    \mathscr{T}_s^{s'}& (\hat{1}_\mathscr{i}^s\otimes\hat{x}_{s'}^s) \mathscr{T}_{s'}^{s},\\
    &=\int x_{s'}\ket{x'_\mathscr{i},x'_{s}}\bra{x'_\mathscr{i},x'_{s}}^{s'}dx'_\mathscr{i} dx'_{s},\\
    &=-\hat{1}_\mathscr{i}^{s'}\otimes\hat{x}^{s'}_s,\\
\end{split}
\end{equation}
and
\begin{equation}
\begin{split}
    \mathscr{T}_s^{s'}& (\hat{1}_\mathscr{i}^s\otimes\hat{p}_{s'}^{s}) \mathscr{T}_{s'}^{s},\\
    &=\int p_{s'}\ket{p'_\mathscr{i},p'_{s}}\bra{p'_\mathscr{i},p'_{s}}^{s'}dp'_\mathscr{i} dp'_{s},\\
    &=-\frac{m_{s'}}{m_s}
\hat{1}_\mathscr{i}^{s'}\otimes\hat{p}_{s}^{s'}.\\
\end{split}
\end{equation}

\section{Compatibility of new quantization rules with $\mathscr{T}_s^{s'}$}
Here, we show that the new quantization rules are compatible with $\mathscr{T}$, while the standard quantization breaks the extended relativity principle.  To prove this, consider technical variables that are also employed in the main text, such as:
\begin{enumerate}[leftmargin=11pt,label=\roman*.]
    \item Canonical Auxiliary Variables: Although the momentum operators are non-canonical, canonical momentum may be introduced as
\begin{equation}
\hat\pi^s_\mathscr{i}
= \hat p^s_\mathscr{i}
-\frac{m_\mathscr{i}}{m_s+\sum_\mathscr{k} m_\mathscr{k}}\sum_\mathscr{k} \hat p^s_\mathscr{k},
\end{equation}
satisfying
\(
[\hat x_\mathscr{i}^s,\hat \pi_\mathscr{j}^s]=i\hbar\delta_{\mathscr{i}\mathscr{j}}.
\) Similarly, one can define the same variables with respect to $s'$.

\item Position representation of the momentum state: 
    The momentum eigenstates in the $s$-frame $\ket{p_\mathscr{i},p_{s'}}^{s}$ is defined by
\begin{equation}
\tilde\Psi^s_p=\braket{x_\mathscr{i},x_{s'}|p_\mathscr{i},p_{s'}}^{s}
,\hspace{0.3cm}
\tilde\Psi^s_p
=\exp\!\left[
\frac{i}{\hbar}(\rho x_\mathscr{i}+\sigma x_{s'})
\right],
\end{equation}
with the eigenvalue equations:
\begin{equation}
\hat{p}_\mathscr{i}^s \tilde\Psi^s_p=p_\mathscr{i}\tilde\Psi^s_p,
\qquad
\hat{p}_{s'}^s\tilde\Psi^s_p=p_{s'}\tilde\Psi^s_p.
\end{equation}
By using Eq.~\eqref{newmomentums}, we get
\begin{equation}
\begin{split}
\left(1+\frac{m_\mathscr{i}}{m_s}\right)\rho
    +\frac{m_\mathscr{i}}{m_s}\sigma
    &=p_\mathscr{i},\\
\frac{m_{s'}}{m_s}\rho
    +\left(1+\frac{m_{s'}}{m_s}\right)\sigma
    &=p_{s'},
\end{split}
\end{equation}
which can be solved by the canonical variables $\rho=\pi_\mathscr{i}^s$ and $\sigma=\pi_{s'}^s$. Then, combining these results, the position representation of the momentum states is
\begin{equation}
\ket{p_\mathscr{i},p_{s'}}^{s}
=\int
e^{\frac{i}{\hbar}
(\pi_\mathscr{i}^s x_\mathscr{i}+\pi_{s'}^s x_{s'})}
\ket{x_\mathscr{i},x_{s'}}^{s}
\,dx_\mathscr{i}\,dx_{s'}.
\end{equation}

\item Proof: Now, we can prove the last statement. For this, consider the general observer transformation  $\mathscr{T}_s^{s'}$ acting on the momentum state  in the position representation
\begin{equation}
\begin{split}
\mathscr{T}_s^{s'}
\ket{p_\mathscr{i},p_{s'}}^{s}
=&\int
e^{\frac{i}{\hbar}
(\pi_\mathscr{i}^s x_\mathscr{i}+\pi_{s'}^s x_{s'})}
\ket{x'_\mathscr{i},x'_{s}}^{s'}dx'_\mathscr{i}dx'_{s},
\end{split}
\end{equation}
then, by replacing $x_\mathscr{i}=x'_\mathscr{i}-x'_s$ and $x_{s'}=-x'_s$, we get
\begin{equation}
\mathscr{T}_s^{s'}
\ket{p_\mathscr{i},p_{s'}}^{s}
=\int
e^{\frac{i}{\hbar}
(\pi_\mathscr{i}^s x'_\mathscr{i}-(\pi_\mathscr{i}^s+\pi_{s'}^s)x'_s)}
\ket{x'_\mathscr{i},x'_{s}}^{s'}dx'_\mathscr{i}dx'_{s}.
\end{equation}
Therefore, the auxiliary canonical momentum transforms as ${\pi'}_\mathscr{i}^{s'}=\pi_\mathscr{i}^s$, and ${\pi'}_s^{s'}=-(\pi_\mathscr{i}^s+\pi_{s'}^s)$, which results in  
\begin{equation}
\mathscr{T}_s^{s'}
\ket{p_\mathscr{i},p_{s'}}^{s}
=
\ket{
p_\mathscr{i}-\frac{m_\mathscr{i}}{m_{s'}}p_{s'},
-\frac{m_s}{m_{s'}}p_{s'}
}^{s'}.
\end{equation}
This shows that the same coordinate transformation uniquely fixes the momentum transformation. Any alternative choice of the canonical momentum, e.g. $\pi\rightarrow p$, will result in another transformation $\ket{
p_\mathscr{i}, p_\mathscr{i}- p_s^{s'} }$, incompatible with the definition of $\mathscr{T}_s^{s'}$.
\end{enumerate}

\section{Covariance of the new quantization rules}
Here, we show the covariance of the quantization rules under general observer transformations. Using the new prescription of Eqs.~\eqref{[x,p]general}, we get
\begin{equation}
\begin{split}
&[\hat{x}_\mathscr{i}^s,\hat{p}_\mathscr{i}^s]
    = i\hbar\!\left(1+\frac{m_\mathscr{i}}{m_s}\right),\\
&[\hat{x}_{s'}^s,\hat{p}_{s'}^s]
    = i\hbar\!\left(1+\frac{m_{s'}}{m_s}\right),\\
[\hat{x}_\mathscr{i}^s,\hat{p}_{s'}^s]
    &= i\hbar\,\frac{m_{s'}}{m_s},\quad [\hat{x}_{s'}^s,\hat{p}_\mathscr{i}^s]= i\hbar\,\frac{m_\mathscr{i}}{m_s}.\\
\end{split}
\end{equation}
Therefore,
\begin{equation}
\begin{split}
[\hat{x}_\mathscr{i}^{s'}&,\hat{p}_\mathscr{i}^{s'}]\\
&= [\mathscr{T}_{s}^{s'}(\hat{x}_\mathscr{i}^{s}-\hat{x}_{s'}^{s})\mathscr{T}_{s'}^{s},\mathscr{T}_{s}^{s'}\big(\hat{p}_\mathscr{i}^{s}-\frac{m_\mathscr{i}}{m_{s'}}\hat{p}_{s'}^{s}\big)\mathscr{T}_{s'}^{s}],\\
&=[\hat x_\mathscr{i}^s,\hat{p}_\mathscr{i}^s]-\frac{m_\mathscr{i}}{m_{s'}}[\hat x_\mathscr{i}^s,\hat{p}_{s'}^s]-[\hat x_{s'}^s,\hat{p}_\mathscr{i}^s]+\frac{m_\mathscr{i}}{m_{s'}}[\hat x_{s'}^s,\hat{p}_{s'}^s],\\
&=i\hbar\bigg(1+\frac{m_\mathscr{i}}{m_s} -\frac{m_\mathscr{i}m_{s'}}{m_{s'}m_s}-\frac{m_\mathscr{i}}{m_s}+\frac{m_\mathscr{i}}{m_{s'}}+\frac{m_\mathscr{i}m_{s'}}{m_{s'}m_s}\bigg),\\
&=i\hbar\bigg(1+\frac{m_\mathscr{i}}{m_{s'}} \bigg),
\end{split}
\end{equation}
also,
\begin{equation}
\begin{split}
[\hat{x}_s^{s'}&,\hat{p}_\mathscr{i}^{s'}]\\
&=[\mathscr{T}_{s}^{s'}(-\hat{x}_{s'}^{s})\mathscr{T}_{s'}^{s},\mathscr{T}_{s}^{s'}\big(\hat{p}_{\mathscr{i}}^{s}-\frac{m_\mathscr{i}}{m_{s'}}\hat{p}_{s'}^{s}\big)\mathscr{T}_{s'}^{s}],\\
&=-[\hat x_{s'}^s,\hat{p}_{\mathscr{i}}^s]+\frac{m_\mathscr{i}}{m_{s'}}[\hat x_{s'}^s,\hat{p}_{s'}^s],\\
&=i\hbar\bigg(-\frac{m_\mathscr{i}}{m_s}+\frac{m_\mathscr{i}}{m_{s'}}+\frac{m_\mathscr{i}m_{s'}}{m_{s'}m_s}\bigg),\\
&=i\hbar\frac{m_\mathscr{i}}{m_{s'}},
\end{split}
\end{equation}
and, similarly
\begin{equation}
\begin{split}
[\hat{x}_\mathscr{i}^{s'}&,\hat{p}_s^{s'}]\\
&= [\mathscr{T}_{s}^{s'}(\hat{x}_\mathscr{i}^{s}-\hat{x}_{s'}^{s})\mathscr{T}_{s'}^{s},\mathscr{T}_{s}^{s'}\big(-\frac{m_s}{m_{s'}}\hat{p}_{s'}^{s}\big)\mathscr{T}_{s'}^{s}],\\
&=-\frac{m_s}{m_{s'}}[\hat x_\mathscr{i}^s,\hat{p}_{s'}^s]+\frac{m_s}{m_{s'}}[\hat x_{s'}^s,\hat{p}_{s'}^s],\\
&=i\hbar\bigg(-\frac{m_sm_{s'}}{m_{s'}m_s}+\frac{m_s}{m_{s'}}+\frac{m_sm_{s'}}{m_{s'}m_s}\bigg),\\
&=i\hbar\frac{m_s}{m_{s'}}.
\end{split}
\end{equation}

\newpage

\section{The three-dimensional extension}
Here, we extend the quantization rules for the three-dimensional case, by considering $\{ \hat {\mathscr{x}}_\mathscr{k}, \hat {\mathscr{p}_x}_\mathscr{k} \},\{ \hat {\mathscr{y}}_\mathscr{k}, \hat {\mathscr{p}_y}_\mathscr{k} \},\{ \hat {\mathscr{z}}_\mathscr{k}, \hat {\mathscr{p}_z}_\mathscr{k} \}$ as the position and momentum of the $\mathscr{i}$ and $\mathscr{j}$ objects relative to $s$ observer of mass $m_s$, where the new commutation holds as
\begin{equation}
\label{Eq2:canonicalcommutationrelation}
\begin{split}
    &[\hat {\mathscr{x}}_\mathscr{i}^{ s}, \hat {\mathscr{p}_x}_\mathscr{j}^{ s}] = i \hbar \bigg(\delta_{\mathscr{i}\mathscr{j}}+\frac{m_\mathscr{j}}{m_s}\bigg),\\
    &[\hat {\mathscr{y}}_\mathscr{i}^{s}, \hat {\mathscr{p}_y}_\mathscr{j}^{ s}] = i \hbar \bigg(\delta_{\mathscr{i}\mathscr{j}}+\frac{m_\mathscr{j}}{m_s}\bigg),\\
    &[\hat {\mathscr{z}}_\mathscr{i}^{ s}, \hat {\mathscr{p}_z}_\mathscr{j}^{ s}] = i \hbar \bigg(\delta_{\mathscr{i}\mathscr{j}}+\frac{m_\mathscr{j}}{m_s}\bigg).\\
\end{split}
\end{equation}
Any other combination is zero. Now, introducing relative coordinates and momenta with respect to observer $s'$, according to
\begin{equation}
\label{Eq3:GalileoTransformationfirstcase}
\begin{aligned}
\hat x_\mathscr{k}^{s'} &= \hat {\mathscr{x}}_\mathscr{k}^{ s} - \hat {\mathscr{x}}_{s'}^{ s}, \\
\hat y_\mathscr{k}^{s'} &= \hat {\mathscr{y}}_\mathscr{k}^{ s} - \hat {\mathscr{y}}_{s'}^{ s},\\
\hat z_\mathscr{k}^{s'} &= \hat {\mathscr{z}}_\mathscr{k}^{ s} - \hat {\mathscr{z}}_{s'}^{ s},\\
({{\hat{p}}_x})_\mathscr{k}^{s'} &= \hat {\mathscr{p}_x}_\mathscr{k}^{ s}- \frac{m_\mathscr{k}}{m_{s'}} \hat {\mathscr{p}_x}_{s'}^{ s},\\
({{\hat{p}}_y})_\mathscr{k}^{s'} &= \hat {\mathscr{p}_y}_\mathscr{k}^{ s}- \frac{m_\mathscr{k}}{m_{s'}} \hat {\mathscr{p}_y}_{s'}^{ s},\\
({{\hat{p}}_z})_\mathscr{k}^{s'} &= \hat {\mathscr{p}_z}_\mathscr{k}^{ s}- \frac{m_\mathscr{k}}{m_{s'}} \hat {\mathscr{p}_z}_{s'}^{ s}.
\end{aligned}
\end{equation}
Therefore, one obtains the same properties of the dimensional case, in which the covariance is given by:
\begin{equation}
\label{primivitenewcommutation}
\begin{split}
[\hat x_\mathscr{i}^{s'}, ({{\hat{p}}_x})_\mathscr{j}^{ {s'}}] = i \hbar \left( \delta_{\mathscr{i}\mathscr{j}} + \frac{m_\mathscr{j}}{m_{s'}} \right),\\
[\hat y_\mathscr{i}^{s'}, ({{\hat{p}}_y})_\mathscr{j}^{ {s'}}] = i \hbar \left( \delta_{\mathscr{i}\mathscr{j}} + \frac{m_\mathscr{j}}{m_{s'}} \right),\\
[\hat z_\mathscr{i}^{s'}, ({{\hat{p}}_z})_\mathscr{j}^{ {s'}}] = i \hbar \left( \delta_{\mathscr{i}\mathscr{j}} + \frac{m_\mathscr{j}}{m_{s'}} \right),
\end{split}
\end{equation}
where any other combinations result in zero.

\section{The algebra of angular momentum }
In this framework, the algebra of angular momentum for particle $\mathscr{i}$, relative to the general observer $s$, is modified as
\begin{equation}
\begin{split}
    [\hat L^s_x,\hat L^s_y]&=i\hbar \bigg(1+\frac{m_\mathscr{i}}{m_s}\bigg) \hat L^s_z,\\
    [\hat L^s_y,\hat L^s_z]&=i\hbar \bigg(1+\frac{m_\mathscr{i}}{m_s}\bigg) \hat L^s_x,\\
    [\hat L^s_z,\hat L^s_x]&=i\hbar \bigg(1+\frac{m_\mathscr{i}}{m_s}\bigg) \hat L^s_y.
\end{split}
\end{equation}
This follows from the conventional definition, $\vec{\hat L}^s=\vec{\hat x}^s\times\vec{\hat p}^s$, and the new quantization rules.

\end{document}